\documentclass[12pt]{article}%
\usepackage{amsmath,amssymb,amsthm,amsfonts}
\usepackage{wasysym}
\usepackage{graphicx}
\usepackage[dvipsnames]{xcolor}
\usepackage{stackengine}

\usepackage[colorlinks]{hyperref}
\usepackage{tikz}
\usepackage{stix}
\usepackage{float}
\usepackage[export]{adjustbox}
\usepackage[labelformat=simple]{subcaption}

\usepackage[margin=1in]{geometry}
\usepackage{appendix}
\usepackage[numbers]{natbib}

\renewcommand{\epsilon}{\varepsilon}

\newcommand{\sgn}{\operatorname{sgn}}

\definecolor{red}{rgb}{0.8500, 0.3250, 0.0980}
\definecolor{green}{rgb}{0.4660, 0.6740, 0.1880}
\definecolor{yellow}{rgb}{0.9290, 0.6940, 0.1250}
\definecolor{blue}{rgb}{0, 0.4470, 0.7410}

\newcommand\blfootnote[1]{%
  \begingroup
  \renewcommand\thefootnote{}\footnote{#1}%
  \addtocounter{footnote}{-1}%
  \endgroup
}

\begin{document}

\title{Multi-bounce resonances in the interaction of walking droplets$^\bigwhitestar$}

\author{George Zhang,\thanks{Department of Applied Mathematics, University of Washington} ~~Ivan C. Christov,\thanks{School of Mechanical Engineering, Purdue University}~\thanks{Department of Computer Science, University of Nicosia} ~~Aminur Rahman\footnotemark[1]~\thanks{AI Institute in Dynamic Systems, University of Washington}~\thanks{Correspondence to \url{arahman2@uw.edu}}}

\date{}

\maketitle

\begin{abstract}
Discrete dynamical models of walking droplets (``walkers'') have allowed swift numerical experiments revealing heretofore unobserved quantum statistics and related behaviors in a classical hydrodynamic system. {We present evidence that one such model of walking droplets exhibits the empirically elusive $n$-bounce resonances that are traditionally seen in the scattering of solitary waves governed by covariant nonlinear field theories with polynomial self-interaction. A numerical investigation of the chosen model of interacting walking droplets reveals a fractal structure of resonances in the velocity in--velocity out diagram, much like the usual maps constructed for collisions of solitary waves. We suggest avenues for further theoretical analysis of walker collisions, which may connect this discrete model to the field-theoretic setting, as well as directions towards new experimental realizations $n$-bounce resonances.}
\end{abstract}

\blfootnote{$\bigwhitestar$ This paper is dedicated to the memory of our friend, mentor, and colleague, Denis Blackmore. Sk\r{a}l, \includegraphics[width = 10pt]{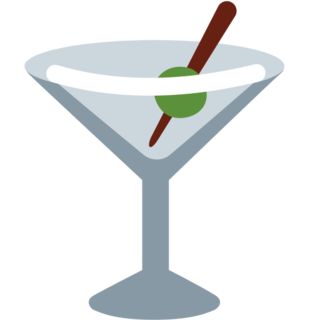}}

\section{Introduction}
\label{sec:intro}


From afar, it may seem like droplets ``walking'' on a vibrating fluid bath and solitary-wave solutions to nonlinear wave equations have little in common. Unlike solitary waves, waves surfacing in the fluid bath, and sustaining the walking droplet, are hardly localized disturbances, nor do they maintain their shape. (Though solitary waves exist on shallow water in a different context \cite{jcr45,kdv95,Dauxois2006}.) With each impact between the droplet and its bath, a new eigenmode is excited in the surface wavefield. Further, a droplet drives the wave generation and propagation whereas, after an initial perturbation, a solitary wave is self-sustained and does not require external energy input to continue propagating---its dynamics being set by the balance of nonlinearity and dispersion in the medium. However, we find that, qualitatively, the dynamics of colliding and interacting walking droplets (``walkers'') are reminiscent of the collisions and interactions of solitary waves \cite{Campbell2019,Goodman2019} under a generic covariant nonlinear field theory, such as the so-called $\phi^4$ one arising in many branches of physics \cite{Manton2004,Vach2006}.

\begin{figure}[hb]
    \centering
    \includegraphics[width = 0.75\textwidth]{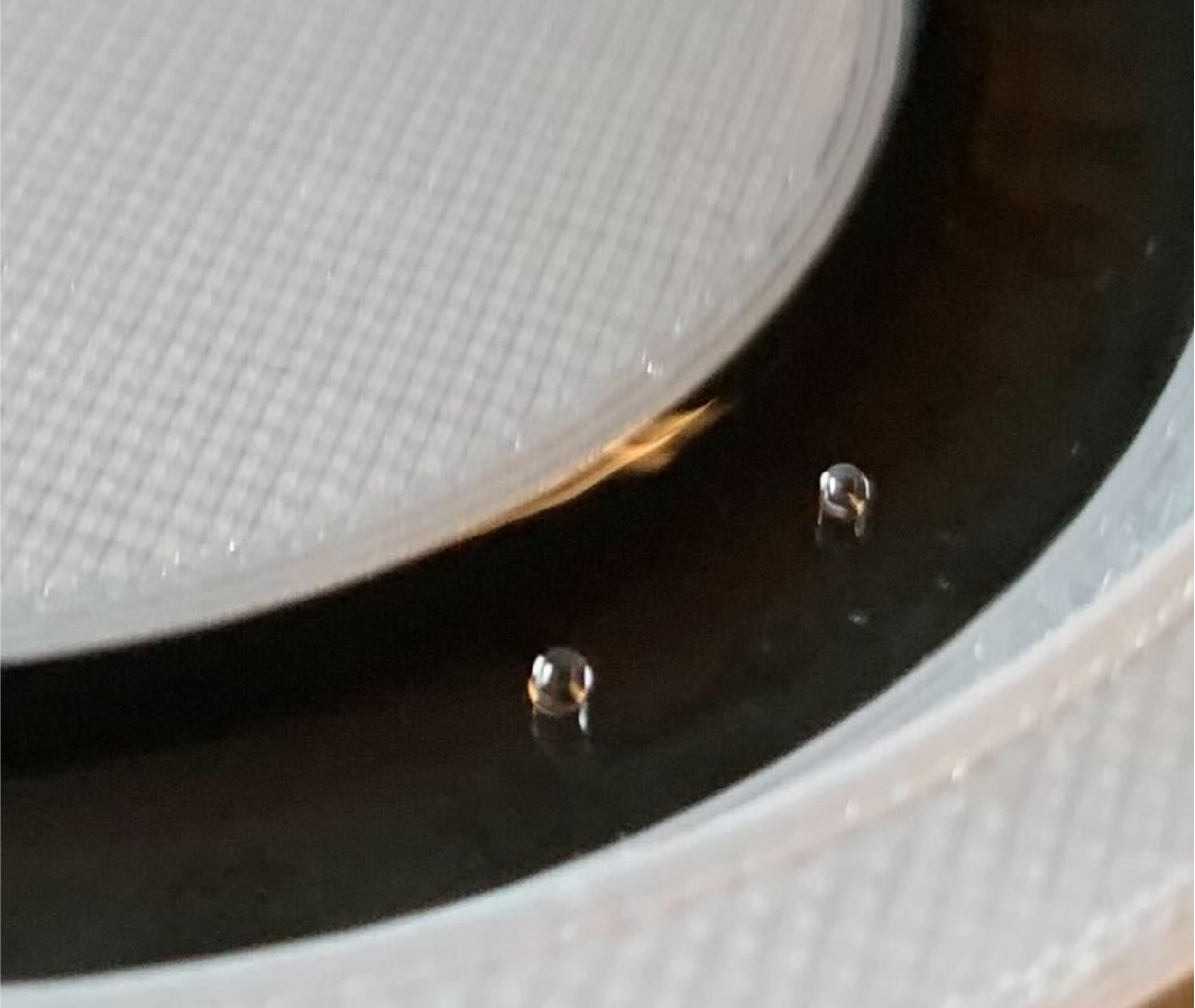}
    \caption{A snapshot of two walking droplets on an annulus interacting with each other through the waves they produce on the fluid bath's surface.}
    \label{Fig: Experimental Picture}
\end{figure}

One of the remarkable aspects of the walker phenomenon is the wave-like statistics arising from the horizontal motion of the droplet, which has been well-delineated in reviews by Bush \cite{Bush10, Bush15a, Bush15b}, Bush and Oza \cite{BushOza2021}, and Rahman and Blackmore \cite{RahmanBlackmoreReview20}.  However, most of the wave-like statistics in both experiments \cite{HMFCB13, Saenz-Mirage2018, RahmanHQAAnalysis2022} and dynamical models \cite{Gilet16, RahmanHQAAnalysis2022, QuintoRahman_Ellipse22} have been shown for a single droplet. One interesting deviation from this comes in the form of a string of droplets exhibiting a traveling wave \cite{ThomsonDurey_Soliton2020, ThomsonCouchman_Soliton2020}.  Between a score or two of droplets are placed around an annulus. As the vertical acceleration of the bath vibration is increased, the droplets start to destabilize, and eventually, a circumferential perturbation from the initial position is transferred through neighboring droplets and travels around the annulus. In this brief communication, we show the existence of solitary wave-like statistics in a discrete dynamical model of the mutual interaction of two walking droplets, as shown in Fig.~\ref{Fig: Experimental Picture}.

{There are a number of detailed modeling and experimental studies on the dynamics and interactions of two walking droplets in the literature. For example, two orbiting droplets were considered in \cite{CPFB05,PBC2006,OSHMB}, two promenading droplets were studied in \cite{Borghesi2014,Arbelaiz2018}, two oscillating droplets were analyzed in \cite{couchman_turton_bush_2019}, two ratcheting droplets were interrogated in \cite{Eddi_2008,G-RCCB}, while scattering of two droplets was examined in \cite{Tadrist2018}. The detailed analysis of the hydrodynamics is not the goal of this brief report. Here, we merely seek to establish the existence and fractal structure of resonances in the velocity in--velocity out diagram for colliding walkers under one simple model of the phenomenon, for which chaotic dynamics were already proved rigorously \cite{Rahman2018}.}

Meanwhile, covariant nonlinear field theories with polynomial self-interaction arise in a number of fields from condensed matter physics \cite{Tinkham1996} to materials science \cite{Gufan1982} to high-energy physics and cosmology \cite{Raja1982,Vach2006}, see also the overview in \cite{sck19}. In the simplest one-dimensional context, these theories take the form of a wave equation (d'Alembertian) forced by a nonlinear (polynomial) function of the dependent variable. These nonlinear wave equations (also referred to as nonlinear Klein--Gordon equations, by a loose analogy to the Klein--Gordon equation arising in quantum field theory) exhibit self-sustained, localized (solitary) traveling wave solutions (so-called topological solitons \cite{Manton2004} or domain walls \cite{Tinkham1996}). These solitary waves can interact by, for example, being sent on a collision course with each other. Perhaps not unexpectedly, the interactions are complex: the solitary waves may repel each other, they may undergo a successive series of collisions (``bounces'') as if they are unable to escape each others' pull, or they may form a permanent bound state \cite{Campbell1983,Anninos1991,Belova1997}. It is hypothesized that these behaviors may be simple models of subatomic particle interactions \cite{Manton2008}. However, the study of collisions of solitary waves in nonlinear field theories remains phenomenological, without direct experimental confirmation (see, e.g., the discussion in \cite{Goodman2015}). In this communication, we show that a strikingly similar collision phenomenology {[compare Fig.~\ref{fig:bounce_windows}(a) to Fig.~\ref{fig:bounce_windows}(b) below] is also observed in a model of the macroscopic, classical system of two walking droplets interacting with each other.}

To this end, we consider a simplified discrete dynamical model of the walker interactions in Sec.~\ref{sec:model}. Then, in Sec.~\ref{sec:collision}, we present observations of $n$-bounce resonances exhibited by the interacting droplets, including the fractal structure of the velocity in--velocity out diagram. Finally, in Sec.~\ref{sec:conclusion} we conclude with a discussion of the connections with solitary wave collisions, and we note future work that could produce physical realizations of the observed effect.


\section{The discrete dynamical model}
\label{sec:model}

In this section, we model the self-propulsion and interaction between two walking droplets as a discrete dynamical system. We start with Eq.~(23) of \cite{Rahman2018}:
\begin{multline}
    v_i(n+1) = C\Bigg[v_i(n) + K \sin\big(\omega v_i(n)\big) e^{-\nu v_i(n)^2} \\
    + K\eta(\gamma n) \sum_{m=1,\,m\ne i}^M \sgn\big( x_i(n) - x_m(n)\big) e^{-\nu\left[x_i(n) - x_m(n)\right]^2}\Bigg],
    \label{eq:v_map}
\end{multline}
where $v_i(n)$ and $x_i(n)$ are the velocity and the position, respectively, of the $i$\textsuperscript{th} droplet (walker) after their $n$\textsuperscript{th} impact with the surface of the fluid reservoir below them, including the mutual effects of up to $M$ other walkers. The `bump function' $\eta$ is defined as
\begin{equation}
    \eta(\xi) = \begin{cases}
        \exp\left(1-\frac{1}{1-\xi^2}\right), &\quad\xi\in(-1,1),\\
        0, &\quad\xi\not\in(-1,1).
    \end{cases}
    \label{Eq: Bump}
\end{equation}
Further, $\gamma\in[0,1]$ and $\nu\in\mathbb{R}^+$ are damping factors, while $K$ and $C$ are parameters modeling the effect of `kicks' on the droplets (walkers).

Equation~\eqref{eq:v_map} is supplemented by the kinematic relation 
\begin{equation}
    x_i(n+1) = x_i(n) + v_i(n+1),
    \label{eq:kinematic}
\end{equation} 
which advances the walker position after the $n$\textsuperscript{th} bounce. 

In \cite{Rahman2018}, a set of interacting droplets converged to a state in which they are equispaced around a circle.  In this study, we wish to model the interaction of two droplets propelled towards each other with equal and opposite velocities: $v_1(0) = v_\text{in}$ and $v_2(0) = -v_\text{in}$.  We know empirically that the droplets repel each other through their respective wavefields, and the strength of the repulsion increases as the droplets get closer to each other. Therefore, we simplify Eq.~\eqref{eq:v_map} as 
\begin{multline}
     v_i(n+1) = C\Bigg[v_i(n) + K \sin\big(\omega v_i(n)\big) e^{-\nu v_i(n)^2} \\
    + K\eta\big(x_i(n) - x_m(n)\big)\sgn\big( x_i(n) - x_m(n)\big) e^{-\nu\left[x_i(n) - x_m(n)\right]^2}\Bigg]
    \label{Eq: Model}
\end{multline}
for the two-walker system in the present work.  

\begin{figure}[b!]
    \centering
    \includegraphics[width = 0.75\textwidth]{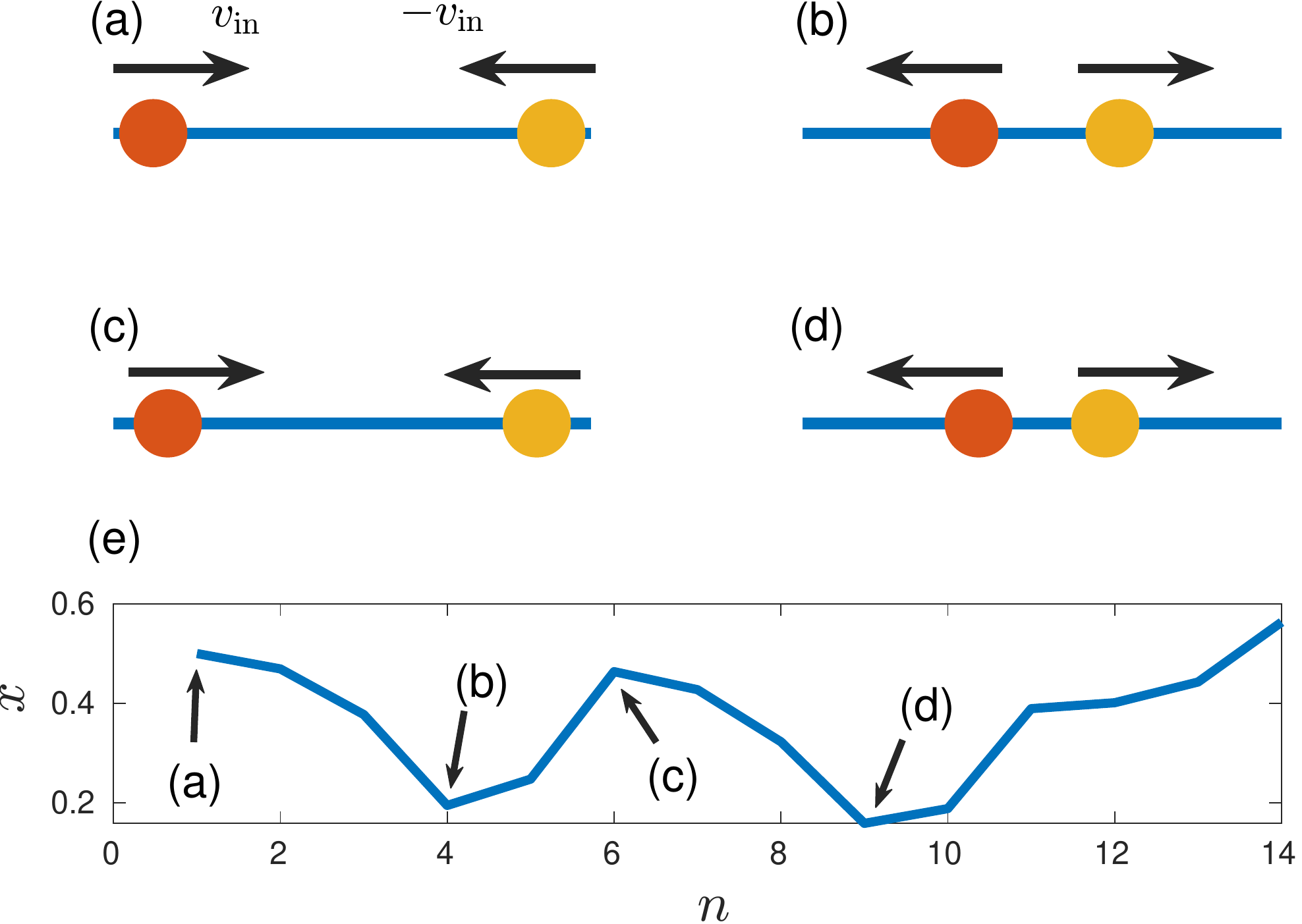}
    \caption{(a-d) Schematic of a pair of walkers undergoing a two-bounce scattering. The walkers (a) approach each other from well-separated initial positions (one could say at effective infinity) with equal but opposite initial velocities, and then (b) they are repelled by their interaction. However, their interaction, coupled with their self-propulsion, (c) brings them back together until (d) they eventually escape each other's domain of influence. (e) The time-discrete trajectory of the left walker, with the events (a-d) labeled on it.}
    \label{fig:schematic}
\end{figure}

An example of the type of behavior produced by the model under these assumptions is shown in Fig.~\ref{fig:schematic}. In this figure, two walking droplets start [Fig.~\ref{fig:schematic}(a)] from well-separated initial condition $-x_1(0) = x_2(0) = 1$ (effective infinity as far as the interaction is concerned; i.e., the boundary of the support of the bump function \eqref{Eq: Bump}).  From these initial points, the droplets are given an initial velocity of $v_1(0) = -v_2(0) = v_\text{in}$.  The droplets approach each other until the repulsion from their interaction overcomes their own inertia, after which they reverse direction [Fig.~\ref{fig:schematic}(b)].  This is known as a \emph{bounce}.  The droplets can bounce once, several times, or be bound forever.  If the droplets are not in a bound state, they eventually escape and go beyond the effective infinity [Fig.~\ref{fig:schematic}(d)].  In this figure, a two-bounce scattering is shown where the droplets move away from each other but then approach one another again [Fig.~\ref{fig:schematic}(c)].  In Fig.~\ref{fig:schematic}(e), we present the complete iterated trajectory of the left droplet for this two-bounce case with the relevant events from Fig.~\ref{fig:schematic}(a-d) labeled on the plot.

\begin{figure}[b!]
    \centering
    \begin{subfigure}[b]{0.45\textwidth}
    \centering
        \includegraphics[width=\columnwidth]{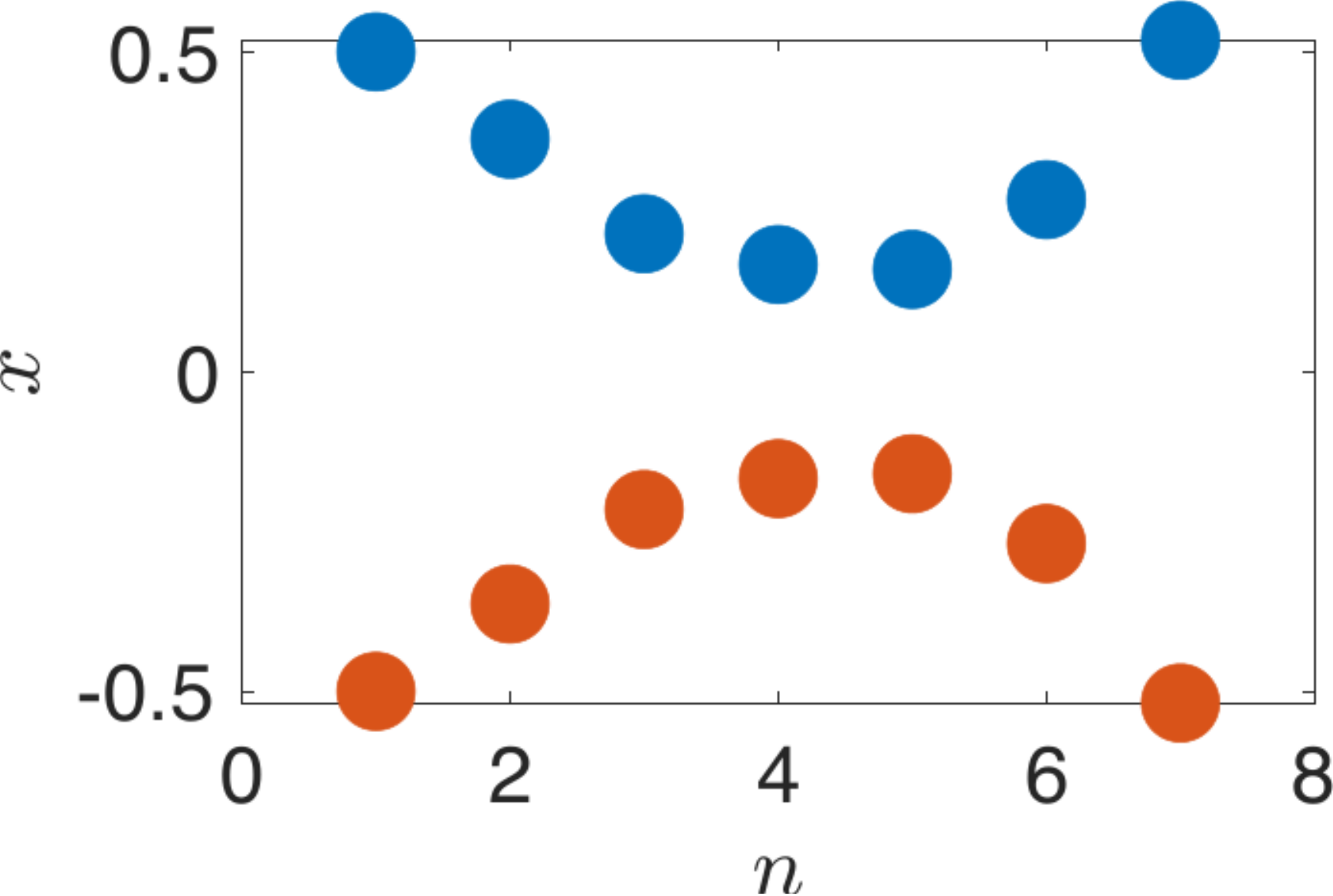}
        \caption{}
    \end{subfigure}
    \begin{subfigure}[b]{0.45\textwidth}
    \centering
        \includegraphics[width=\columnwidth]{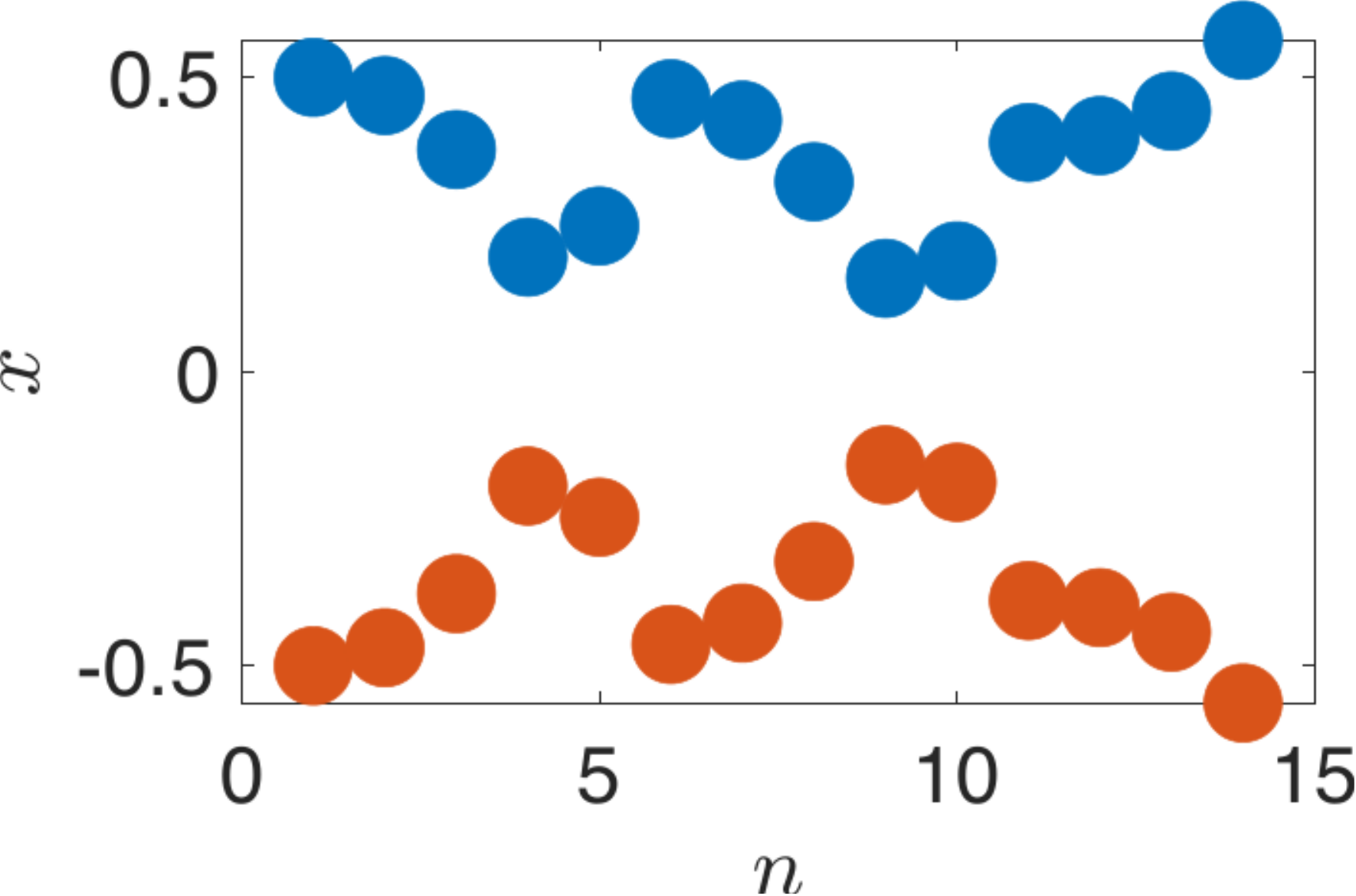}
        \caption{}
    \end{subfigure}
    \\
    \begin{subfigure}[b]{0.45\textwidth}
    \centering
        \includegraphics[width=\columnwidth]{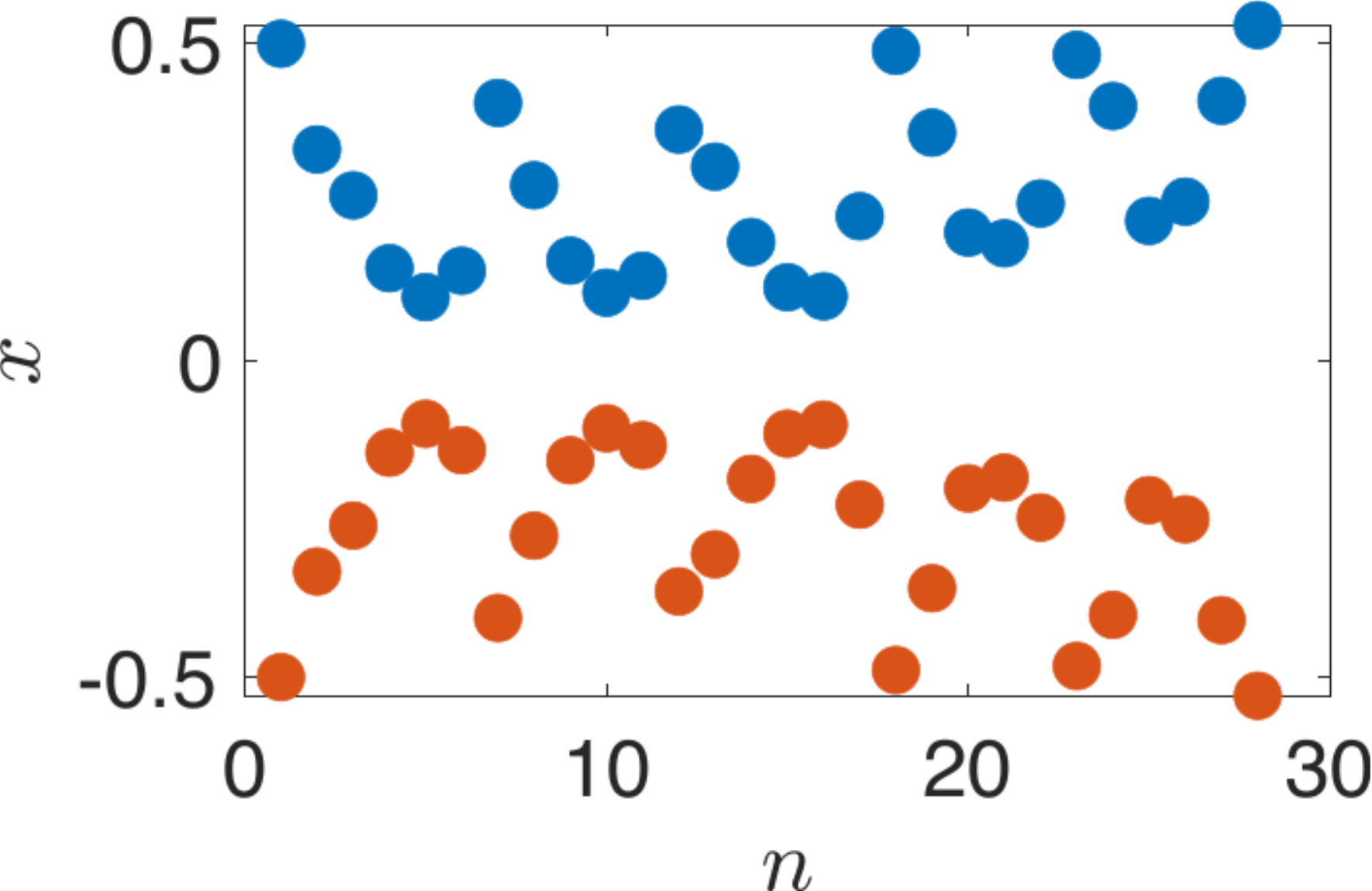}
        \caption{}
    \end{subfigure}
    \begin{subfigure}[b]{0.45\textwidth}
    \centering
        \includegraphics[width=\columnwidth]{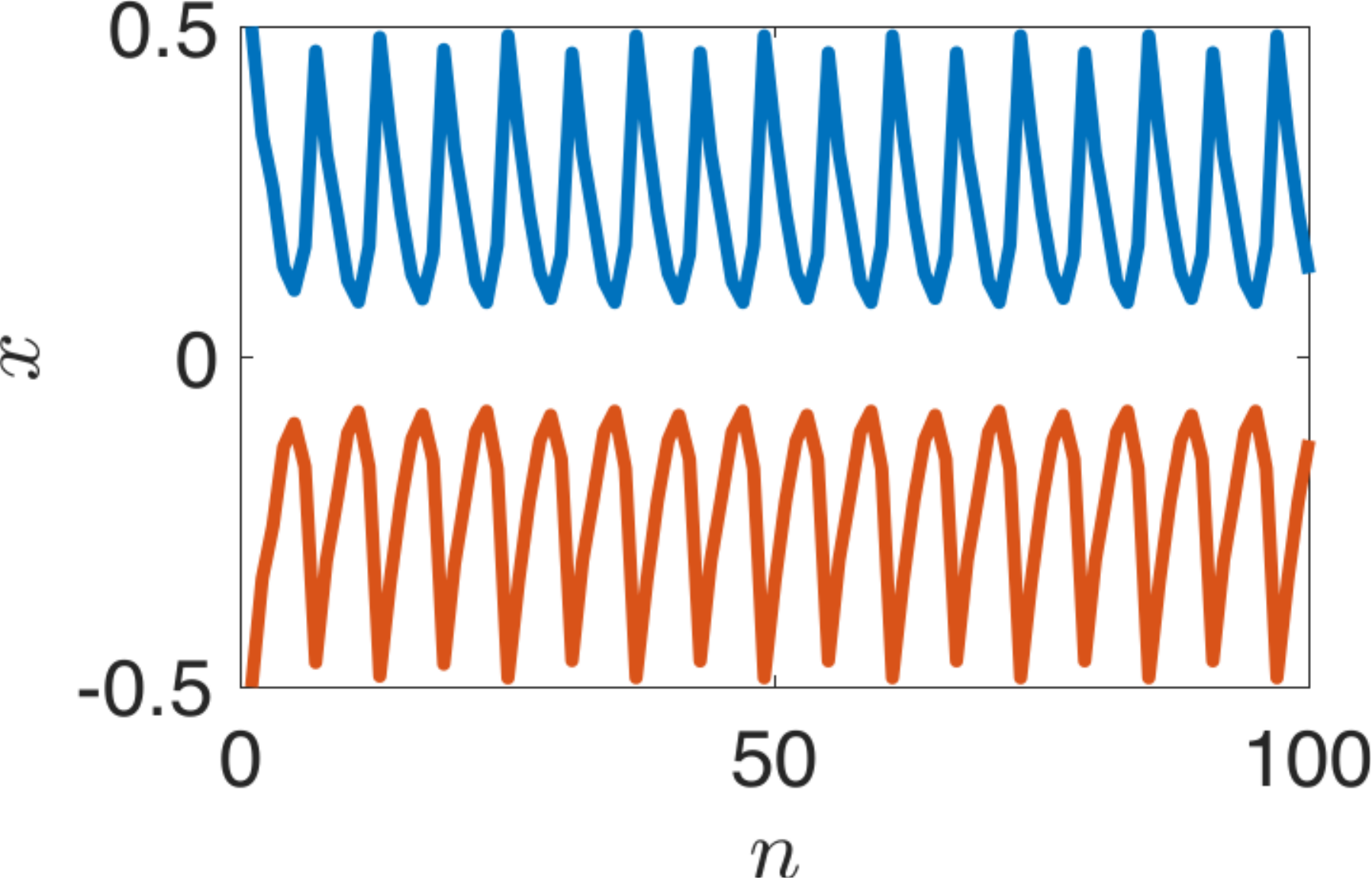}
        \caption{}
    \end{subfigure}  
    \caption{Example discrete walker trajectories exhibiting (a) a one-bounce with $v_\text{in} = 0.15$, (b,c) multi-bounce behavior with $v_\text{in} = 0.01$ and $v_\text{in} = 0.1335045$, respectively, and (d) a bound state with $v_\text{in} = 0.135$.  For each plot we use the standard parameters $\omega = 31/2$, $\nu = \omega^2/8.4\pi^2$, and $K = -\pi e^{\nu\pi^2}/\sin(\pi\omega)$ \cite{Rahman2018}, and $C = 1/5K$.}
    \label{fig:example_bounce}
\end{figure}


\section{Collision phenomenology and multi-bounce windows}
\label{sec:collision}

In this section, we shall demonstrate the droplet interactions as modeled by Eq.~\eqref{Eq: Model}.  Throughout the section we will use $\omega = 31/2$, $\nu = \omega^2/8.4\pi^2$, and $K = -\pi e^{\nu\pi^2}/\sin(\pi\omega)$ \cite{Rahman2018}. We may also vary $C$ to produce varying complexity in the velocity iterate space.  For this section, we set $C = 1/5K$, which strikes a balance between the strength of the interaction and self-propulsion.

To demonstrate the existence of multi-bounce windows in the collision of $M=2$ walkers, consider Eq.~\eqref{Eq: Model} for different initial velocities $v_\text{in}$. In Fig.~\ref{fig:example_bounce}, we present four illustrative cases of the behavior of the model. We first take an initial velocity of $v_\text{in} = 0.15$ in Fig.~\ref{fig:example_bounce}(a). The walkers approach each other but are then repelled with a strong enough force to escape past the support of \eqref{Eq: Bump}. In Fig.~\ref{fig:example_bounce}(b), for $v_\text{in} = 0.01$, we observe two-bounce scattering, which is similar to the dynamics shown in the schematic Fig.~\ref{fig:schematic}.  There are also cases of several bounces before escape, as observed in Fig.~\ref{fig:example_bounce}(c) for the initial velocity of $v_\text{in} = 0.1335045$.  Finally, we observe a bound state, in which the walkers can never escape each other, for $v_\text{in} = 0.135$ in Fig.~\ref{fig:example_bounce}(d).

\begin{figure}
    \centering
    \includegraphics[width=0.8\textwidth]{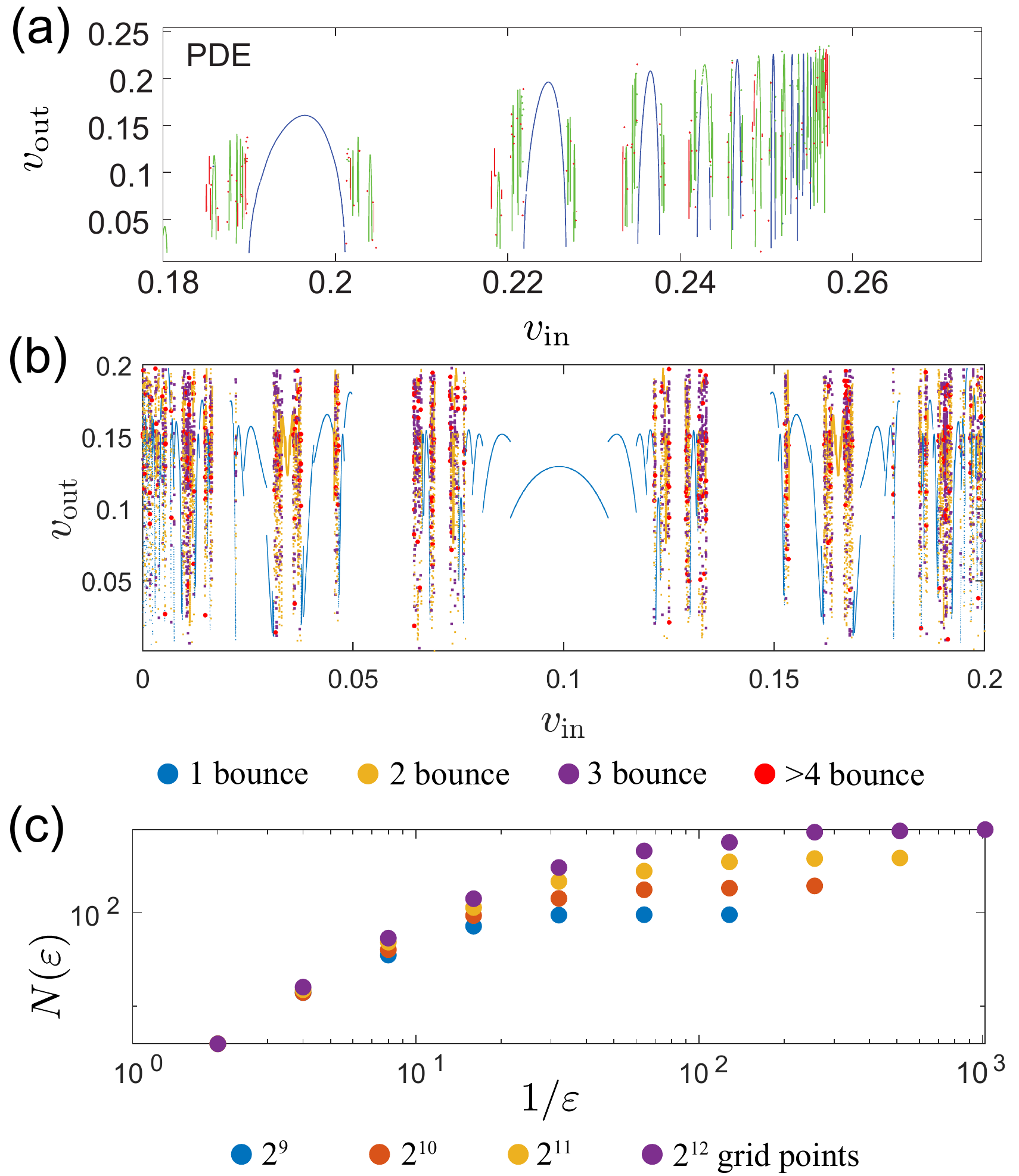}
    \caption{(a) Bounce windows with fractal structure of kink--antikink collisions computed from numerical solutions of the $\phi^4$ nonlinear field equation $\phi_{tt}-\phi_{xx}+\phi-\phi^3=0$ \cite{Goodman2015}. Reprinted from [Goodman, R.H., Rahman, A., Bellanich, M.J., Morrison, C.N., 2015. A mechanical analog of the two-bounce resonance of solitary waves: Modeling and experiment. Chaos 25, 043109], with the permission of AIP Publishing. 
    (b) Bounce windows with fractal structure of the discrete dynamical model of walking droplets considered herein.  The markers are color-coded based on the number of bounces the droplets experienced before escaping.  Blue represents one bounce, yellow represents two bounces, purple represents three bounces, and red represents four or more bounces.  (c) Progression of the number of box coverings required to cover the entire plot (b) to illustrate the calculation of the Minkowski--Bouligand dimension of this fractal set.}
    \label{fig:bounce_windows}
\end{figure}

Constructing a typical velocity in--velocity out diagram colored by the number of bounces the walkers perform seems to exhibit a fractal structure of windows as is the case in kink--anti-kink interactions of the $\phi^4$ nonlinear field theory \cite{Goodman2005,Goodman2019}.  As seen from  Fig.~\ref{fig:bounce_windows}(b), we have quite a range of possible escape velocities $v_\text{out}$ and number of bounces, depending on the initial velocity $v_\text{in}$. {Parts of the diagram show a smooth variation of $v_\text{in}$ with $v_\text{out}$, while other parts of the diagram show unpredictable variation of $v_\text{in}$ and changes in the number of bounces. This kind of diagram is both reminiscent of the corresponding diagram for solitary wave collisions, as in Fig.~\ref{fig:bounce_windows}(a), and also the iterate space produced by map models of these interactions \cite{Goodman2005,Goodman2019}. 

Since we have reason to believe that Fig.~\ref{fig:bounce_windows}(b) has fractal structure}, it is natural to compute the fractal dimension.  We use the Minkowski--Bouligand dimension,
\begin{equation}
    d = \lim_{\varepsilon \rightarrow 0} \frac{\ln(N(\varepsilon))}{\ln(1/\varepsilon)},
    \label{Eq: Fractal}
\end{equation}
with the progression of the number of box coverings, $N(\varepsilon)$ for respective length, $\varepsilon$, of the square covering shown in Fig.~\ref{fig:bounce_windows}(c).  Using this definition gives us a dimension of $d \approx 0.95$. Appendix \ref{Sec: Fractal} provides the details of calculating $d$.

\section{Conclusion}
\label{sec:conclusion}

In this brief communication, we showed that a phenomenological model~\eqref{Eq: Model}, based on the one introduced in \cite{Rahman2018},  exhibits single and multi-bounce behavior and bound states for a pair of walking droplets, much like the behavior often observed in the nonlinear interaction of solitary waves. Further still, we observed apparent fractal structure in the $v_\text{in}$--$v_\text{out}$ diagram for the walking droplet interactions. To further substantiate the latter, we computed the diagram's Minkowski--Bouligand dimension \eqref{Eq: Fractal}.  All of these features are reminiscent of kink--antikink collisions and scattering \cite{Goodman2007, Goodman2005, Anninos1991}, {and Appendices \ref{app:time-cont} and \ref{app:weak-mem} provide further conceptual links between the time-continuous version of the proposed model~\eqref{Eq: Model} and the collective coordinate description of solitary wave interactions. In future work, it would also be of interest to determine whether more refined models of walkers, such as those in \cite{ORB13,DM17}, also lead to single and multi-bounce behavior, bound states, and bounce windows with fractal structure. We expect that they would.}

It would certainly be presumptuous to claim that simulations from a reduced discrete dynamical model indicate that the physical system exhibits $n$-bounce resonances. However, we hope that this investigation may act as a compass for researchers to search for $n$-bounce resonances in the novel physical context of walking droplets and the corresponding experiments, drawing further analogies between phenomenological concepts in nonlinear field theory and dynamics realizable in a bench-top experiment.

\section*{Acknowledgement}
G.Z.\ and A.R.\ appreciate the support of the Department of Applied Mathematics at the University of Washington.
I.C.C.\ would like to acknowledge the hospitality of the University of Nicosia, Cyprus, where this work was completed thanks to a Fulbright U.S.\ Scholar award from the U.S.\ Department of State.



\bibliographystyle{elsarticle-num}

\bibliography{phi4walkers}


\appendix

\section{Appendix}

\subsection{Time-continuous dynamical model}
\label{app:time-cont}

We first wish to convert Eq.~\eqref{eq:v_map} from a discrete-time to a continuous-time dynamical system, under the assumption that $x_i(n+1) - x_i(n)$ and $v_i(n+1) - v_i(n)$ are ``small.'' We follow this approach, which Meiss \cite[p.~806]{meiss1992} employs to turn the (discrete) standard map in the (continuous) equation of motion of a pendulum. Let us first modify the right-hand side of Eq.~\eqref{eq:v_map} so that $v_i$ is not multiplied by $C$. Then, let us further restrict to the case of $M=2$, which turns Eqs.~\eqref{Eq: Model} and \eqref{eq:kinematic} into the system
\begin{subequations}
\begin{align}
    \dot{v}_1(t) &= C K \sin\big(\omega
     {v}_1(t)\big) e^{-\nu {v}_1(t)^2} \label{eq:v1_ode}\\
    &\phantom{=}+ K\eta\big( {x}_1(t) - {x}_2(t)\big) \sgn\big( {x}_1(t) - {x}_2(t)\big)  e^{-\nu\left[{x}_1(t) - {x}_2(t)\right]^2}, \nonumber\\
    \dot{x}_1(t) &= v_1(t),\\
    \dot{v}_2(t) &= C K \sin\big(\omega {v}_2(t)\big) e^{-\nu {v}_2(t)^2}  \label{eq:v2_ode}\\
    &\phantom{=}+ K\eta\big( {x}_2(t) - {x}_1(t)\big) \sgn\big( {x}_2(t) - {x}_1(t)\big)  e^{-\nu\left[{x}_2(t) - {x}_1(t)\right]^2}, \nonumber\\
    \dot{x}_2(t) &= v_2(t),
\end{align}\label{eq:v1_v2_odes}\end{subequations}
where overdots denote derivatives with respect to $t$, and we henceforth leave the time dependence to be understood.

Now, without loss of generality, assume that $x_1<0$ and $x_2>0$ during an interaction. The walkers cannot cross paths due to the repulsive mutual interaction. Hence, Eqs.~\eqref{eq:v1_v2_odes} become
\begin{subequations}
\begin{align}
    \dot{v}_1 &= \hat{K}
 \sin\big(\omega
    {v}_1\big) e^{-\nu {v}_1^2} - K F\big( {x}_2 - {x}_1\big) , \label{eq:v1_ode_2}\\
    \dot{x}_1 &= v_1,\\
    \dot{v}_2 &= \hat{K} \sin\big(\omega {v}_2\big) e^{-\nu {v}_2^2} + K F\big( {x}_2 - {x}_1\big), \label{eq:v2_ode_2}\\
    \dot{x}_2 &= v_2,
\end{align}\label{eq:v1_v2_odes_2}\end{subequations}
having defined $\hat{K}:=CK$ for convenience and the mutual interaction force as
\begin{equation}
    F(\xi) := \eta(\xi) e^{-\nu\xi^2}.
    \label{eq:force}
\end{equation}
Eliminating the $v_i$ between Eqs.~\eqref{eq:v1_v2_odes_2}, we obtain 
\begin{subequations}
\begin{align}
    \dot{x}_1 &= \hat{K} \sin\big(\omega\dot{x}_1\big) e^{-\nu \dot{x}_1^2}  
     - K F\big( {x}_2 - {x}_1\big),
    \label{eq:th1_ode}\\
    \ddot{x}_2 &= \hat{K} \sin\big(\omega\dot{x}_2\big) e^{-\nu \dot{x}_2^2}  
    + K F\big( {x}_2 - {x}_1\big).
    \label{eq:th2_ode}
\end{align}\label{eq:th1_th2_odes}\end{subequations}
Equations~\eqref{eq:th1_th2_odes} bear some similarity to those discussed in \cite{Goodman2007} (see also \cite{Goodman2005,Anninos1991}).

\subsection{Weak-memory limit and effective Lagrangian}
\label{app:weak-mem}

Note that the walker's self-interaction memory (i.e., the sine and exponentials on the right-hand sides of Eqs.~\eqref{eq:th1_th2_odes}) depends on $v_i=\dot{x}_i$. Making a small slope approximation and neglecting higher-order terms, we obtain
\begin{subequations}
\begin{align}
    \dot{x}_1 &= \hat{\hat{K}} \dot{x}_1   
     - K F\big( {x}_2 - {x}_1\big),
    \label{eq:th1_ode_2}\\
    \ddot{x}_2 &= \hat{\hat{K}} \dot{x}_2 
    + K F\big( {x}_2 - {x}_1\big).
    \label{eq:th2_ode_2}
\end{align}\label{eq:th1_th2_odes_2}\end{subequations}
having defined $\hat{\hat{K}}:=CK\omega$ for convenience.

\begin{figure}
    \centering
    \includegraphics[width=0.75\textwidth]{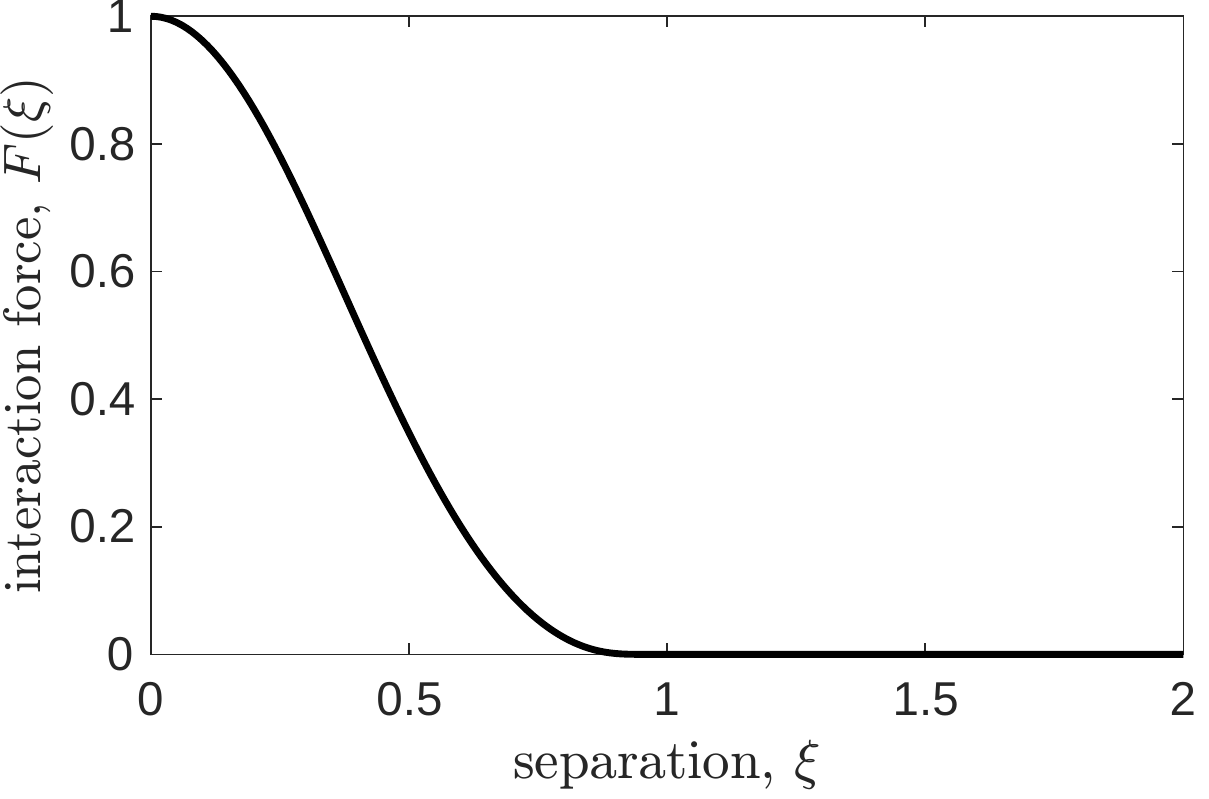}
    \caption{Interaction force between two walkers in the reduced time-continuous dynamical description, as a function of their separation $\xi = x_1 - x_2$, with $\omega=15.5$ and $\nu\approx2.8979$.}
    \label{fig:interaction_potential}
\end{figure}

Now, we observe that Eqs.~\eqref{eq:th1_th2_odes_2} above are a system of Hamiltonian equations for the trajectories $\{x_i(t)\}$, derivable from the following \emph{effective} Lagrangian density:
\begin{equation}
    \mathcal{L}_\mathrm{eff}[x_1(t),x_2(t)] = \frac{1}{2}\dot{x}_1^2 + \frac{\hat{\hat{K}}}{2}x_1^2
    + \frac{1}{2}\dot{x}_2^2 + \frac{\hat{\hat{K}}}{2}x_2^2 
    - K U( {x}_2 - {x}_1),
    \label{eq:Lagrangian}
\end{equation}
by extremizing the action $\int 
\mathcal{L}_\mathrm{eff}[x_1(t),x_2(t)] \,\mathrm{d}t$. In Eq.~\eqref{eq:Lagrangian}, $U(\xi)$ is the force potential such that $-U'(\xi) = F(\xi)$. Further, observe that Eq.~\eqref{eq:Lagrangian} is also the Lagrangian of two nonlinearly coupled oscillators. The interaction force $F(\xi)$ is shown in Fig.~\ref{fig:interaction_potential}. Clearly, the force is repulsive and monotone, which explains why walkers scatter (bounce) rather than annihilate or pass through each other. A more detailed analysis would be needed to reveal the multiple-bounce phenomena reported above.

Interestingly, due to the existence of a Lagrangian for the ODE system~\eqref{eq:th1_th2_odes_2}, this description of the interaction of two walkers can be thought of as a type of \emph{collective coordinate} approximation \cite{Dauxois2006} (see also \cite{Sugiyama1979,Campbell1983,Belova1997,Malomed2002}) commonly used to produce reduced-order models to study the interaction of solitary waves governed by nonlinear field theories. This analogy provides a framework for further analytical progress.

\subsection{Calculating the fractal dimension}
\label{Sec: Fractal}

We calculate the Minkowski--Bouligand dimension by using the box-counting procedure.  We use $0.2/2^{p-3}$ grid points, where $p = 9, 10, 11, 12$, for $v_\text{in}$.  In each case $v_\text{in}$ is partitioned into $2^j$ intervals of equal size, which corresponds to partitioning the square $(0,0.2)\times(0,0.2)$ into $2^j$ rectangles.  Using $0.2/2^{p-3}$ points for each interval in the $v_\text{in}$ direction, we sum up the number of intervals in the $v_\text{out}$ direction of length $0.2/2^j$ that contains a $v_\text{out}$ in each rectangle.  This effectively partitions $v_\text{out}$ into $2^j$ intervals, which then partitions the entire space into $2^{2j}$ squares (or boxes).  We observe that for each case of $p = 9,\ldots, 12$, the Minkowski--Bouligand dimension defined via Eq.~\eqref{Eq: Fractal} gives the same approximation, namely $d \approx 0.95$.

\end{document}